\documentclass[12pt]{article}\usepackage{kakie}
\begin{document}

\title{An Introduction to\\
Cartan's KAK Decomposition\\
for QC Programmers }

\author{Robert R. Tucci\\
        P.O. Box 226\\
        Bedford,  MA   01730\\
        tucci@ar-tiste.com}

\date{ \today}

\maketitle

\vskip2cm
\section*{Abstract}
This paper presents no
new results; its
goals are purely pedagogical.
A special case of the
Cartan Decomposition has found
much utility in the field
of  quantum computing, especially
in its sub-field of
quantum compiling.
This special case
allows one to factor
a general 2-qubit operation
(i.e., an element of $U(4)$)
into local operations
applied
before and after a
three parameter,
non-local operation.
In this paper, we give a
complete and rigorous
proof of this special case of
Cartan's Decomposition.
From the point of view
of QC programmers who might
not be familiar with the
subtleties of
Lie Group Theory,
the proof given here
has the virtues,
that
it is constructive
in nature, and that it
 uses only Linear Algebra.
The constructive proof
presented in this paper is implemented
in some Octave/Matlab
m-files that are included with the paper.
Thus, this paper
serves as documentation for
the attached m-files.

\section{Introduction and Motivation}

Cartan's KAK Decomposition
was discovered by the awesome
mathematical genius, Elie Cartan (1869-1951).
Henceforth, for succinctness,  we
will  refer to
his decomposition merely as KAK.
The letters KAK come from the fact that
in stating and proving KAK, one
considers a group
$\ul{G}=\exp(\ul{g})$
with a subgroup $\ul{K}=\exp(\ul{k})$
and a Cartan subalgebra $\ul{a}$,
where $\ul{g} = \ul{k}\oplus \ul{k}^\perp$
and $\ul{a}\subset \ul{k}^\perp$.
Then one shows that any
$G\in \ul{G}$
can be expressed as $G=K_1 A K_2$, where
$K_1, K_2\in \ul{K}$ and $A\in \exp(\ul{a})$.
An
authoritative discussion of KAK can be
found in the book by Helgason\cite{Helg}.

KAK was first applied to
quantum computing (QC) by
Khaneja and Glaser
in Refs.\cite{KG}.
Since we are using
``KAK" to refer to the general theorem,
we will use ``KAK1" to refer to
the special case
of KAK used by Khaneja and Glaser.
Besides KAK1, the
Cosine-Sine Decomposition (CSD)\cite{Golub}\cite{Paige}
 is another decomposition that
is very useful\cite{Rudi}
in QC.
After Refs.\cite{KG} and \cite{Rudi},
QC workers came to
the  realization\cite{SS-Gestapo}
that CSD
also follows from KAK ,
even
though CSD was discovered\cite{Paige} quite
independently from KAK.

This paper will only discuss KAK1.
KAK1 is the assertion that:
Given any $U\in SU(4)$,
one can find
$A_1, A_0, B_1, B_0\in SU(2)$ and
$\vec{k}\in \RR^3$
so that

\beq
U= (A_1\otimes A_0)\eiks (B_1\otimes B_0)
\;,
\eeq
where $\vec{\Sigma}$ is an operator
that is independent of $U$
and will be defined later.
Thus KAK1 parameterizes $SU(4)$,
a 15-parameter Lie Group,
so that 12 parameters
characterize
local operations, and only 3 parameters
(the 3 components of $\vec{k}$)
characterize non-local ones.

Ever since
 Refs.\cite{KG} appeared,
many
workers other than Khaneja
and Glaser
have used KAK1 in QC
to great advantage
(see, for example,
Refs.\cite{VD}, \cite{VW}, \cite{Zhang}).
Mainly, they have used KAK1 to
compile 2-qubit operations.
For instance,
Vidal and Dawson used KAK1 to
prove that
any 2-qubit
operation can be expressed
with 3 or fewer CNOTs
and some 1-qubit rotations.

This paper includes
a complete, rigorous proof of KAK1
and related theorems.
The proof of KAK1 presented here
is based on the well known
isomorphism $SO(4) =
\frac{SU(2)\times SU(2)}{\{(1,1),(-1-1)\}}$
and on a theorem
by Eckart and Young (EY)\cite{EY}.
The EY theorem gives necessary
and sufficient conditions
for simultaneous SVD (singular
value decomposition) of two matrices.
The relevance of the EY theorem
to KAK1 was pointed out in
Ref.\cite{Gsponer}.
The
proof of KAK1 given here
is a constructive proof, and
it uses only Linear Algebra.
Contrast this to the proof of KAK  given
in Ref.\cite{Helg}, which, although
much more general, is a
non-constructive (``existence") proof,
and it uses advanced concepts in
Lie Group Theory.

Octave is
a programming environment and language
that is gratis and open software. It
copies most of Matlab's function names
and capabilities in
Linear Algebra.
A collection of Octave/Matlab
m-files that implement the algorithms
in this
paper,
can be found at ArXiv (under the
``source" for this paper),
and at my website (www.ar-tiste.com).

\section{Notation and Other Preliminaries}
In this section, we will
define some notation that is
used throughout this paper.
For additional information about our
notation, see Ref.\cite{Paulinesia}.

We will use the word
``ditto" to mean likewise
and respectively. For example,
``$x$ (ditto, $y$)
is in $A$ (ditto, $B$)",
means $x$ is in $A$ and $y$ is in $B$.

As usual, $\RR, \CC$
will stand for the real and complex numbers.
For any complex matrix $A$, the symbols
$A^*, A^T, A^\dagger$ will stand
for the complex conjugate, transpose,
and Hermitian conjugate, respectively,
of $A$.
(Hermitian conjugate
a.k.a. conjugate transpose and
adjoint)

The Pauli matrices are defined by:

\beq
\sigx=
\left(
\begin{array}{cc}
0 & 1\\
1 & 0
\end{array}
\right)
\;,
\;\;
\sigy=
\left(
\begin{array}{cc}
0 & -i\\
i & 0
\end{array}
\right)
\;,
\;\;
\sigz=
\left(
\begin{array}{cc}
1 & 0\\
0 & -1
\end{array}
\right)
\;.
\eeq
They satisfy
\beq
\sigx \sigy = -\sigy \sigx = i \sigz
\;,
\eeq
and the two other equations
obtained from this one by permuting
the indices $(x,y,z)$ cyclically.
We will also have occasion
to use the operator $\vec{\sigma}$, defined by:

\beq
\vec{\sigma} = (\sigx, \sigy, \sigz)
\;.
\eeq

Let $\sigma_{X_\mu}$
for $\mu \in\{0,1,2,3\}$
be defined by
$\sigma_{X_0}=\sigma_1 = I_2$,
where $I_2$ is the 2 dimensional identity matrix,
$\sigma_{X_1}=\sigx$,
$\sigma_{X_2}=\sigy$,
and
$\sigma_{X_3}=\sigz$.
Now define

\beq
\sigma_{X_\mu X_\nu}=
\sigma_{X_\mu}\otimes \sigma_{X_\nu}
\;
\eeq
for $\mu, \nu\in\{0,1,2,3\}$.
For example, $\sigxy= \sigx\otimes \sigy$
and $\sigux = I_2\otimes\sigx$.
The matrices $\sigma_{X_\mu X_\nu}$
satisfy
\beq
\sigxx \sigyy = \sigyy \sigxx = -\sigzz
\;,
\label{eq-com-sigxx-sigyy}
\eeq
and
the two other equations
obtained from this one by permuting
the indices $(x,y,z)$ cyclically.
We will also have occasion
to use the operator $\vec{\Sigma}$, defined by:

\beq
\vec{\Sigma} = (\sigxx, \sigyy, \sigzz)
\;.
\eeq

Define

\beq
\unor=
\frac{1}{\sqrt{2}}
\left(
\begin{array}{cccc}
 1 & 0 & 0 & i\\
 0 & i & 1 & 0\\
 0 & i &-1 & 0\\
 1 & 0 & 0 &-i
\end{array}
\right)
\;.
\eeq
It is easy to check that
$\unor$ is a unitary matrix.
The columns of
$\unor$ are an orthonormal basis,
often called the ``magic basis" in the
quantum computing literature. (That's
why we have chosen to
call this matrix $\unor$,
because of the ``m" in magic).

In this paper, we often
need to find the outcome
$\unor^\dagger X \unor$
(or $\unor X \unor^\dagger$) of
a similarity transformation (
equivalent to a change of
basis) of a matrix $X\in \CC^{4\times 4}$ with
respect to $\unor$. Since $X$ can
always be expressed as a linear
combination of the $\sigma_{X_\mu X_\nu}$,
it is useful to know
the outcomes
$\unor^\dagger(\sigma_{X_\mu X_\nu})\unor$
(or $\unor(\sigma_{X_\mu X_\nu})\unor^\dagger$)
for $\mu, \nu\in\{0,1,2,3\}$. One finds
the following two tables:

\beq
\unor^\dagger(A\otimes B)\unor=\;\;\;\;
\begin{array}{|rr|rrrr|}\hline
&&& B &\rarrow&\\
            &       &1       & \sigx  & \sigy  & \sigz\\ \hline
            &1      &1       &-\siguy & \sigyz &-\sigyx\\
A           &\sigx  &-\sigzy & \sigzu &-\sigxx &-\sigxz\\
\downarrow  &\sigy  &-\sigyu & \sigyy &-\siguz & \sigux\\
            &\sigz  &-\sigxy & \sigxu & \sigzx & \sigzz\\\hline
\end{array}
\;,\label{tab-mh-m}
\eeq

\beq
\unor(A\otimes B)\unor^\dagger=\;\;\;\;
\begin{array}{|rr|rrrr|}\hline
&&& B &\rarrow&\\
            &       &1       & \sigx  & \sigy  & \sigz\\ \hline
            &1      &1       & \sigyz &-\sigux &-\sigyy\\
A           &\sigx  & \sigzx &-\sigxy &-\sigzu &-\sigxz\\
\downarrow  &\sigy  &-\sigyu &-\siguz & \sigyx & \siguy\\
            &\sigz  & \sigxx & \sigzy &-\sigxu & \sigzz\\\hline
\end{array}
\;.\label{tab-m-mh}
\eeq

\section{Proof of KAK1}
In this section, we present
a proof of KAK1 and related theorems.
The proofs
are constructive in nature
and yield
the algorithms used in our software
for calculating KAK1.
Thus, even those persons that are not
too enamored with
mathematical proofs may
benefit from  reading this section.

\begin{theo}\label{claim-isom3}
Define a map $\phi$ by

\beq
\phi: SU(2)\rarrow SO(3)\;,\;\;
\phi(A)= \unor^\dagger (A\otimes A^*) \unor
\;.
\eeq
Then $\phi$ is a well defined, onto,
 2-1, homomorphism.
 Well-defined: For all $A\in SU(2)$,
$\unor^\dagger (A \otimes A^*)\unor \in SO(3)$.
Onto:
For all $Q\in SO(3)$,
 there exist $A \in SU(2)$ such that
$Q= \unor^\dagger (A \otimes A^*)\unor$.
2-1: $\phi$ maps exactly two
elements ($A$ and $-A$) into one ($\phi(A)$).
Homomorphism: $\phi$
preserves group operations.
\end{theo}

\begin{theo}\label{claim-isom4}
Define a map $\Phi$ by

\beq
\Phi: SU(2)\times SU(2)\rarrow SO(4)
\;,\;\;
\Phi(A, B)= \unor^\dagger (A\otimes B^*)\unor
\;.
\eeq
Then $\Phi$ is a well defined, onto,
 2-1,  homomorphism.
Well-defined: For all $
A, B\in SU(2)$,
$\unor^\dagger (A \otimes B^*)\unor \in SO(4)$.
Onto: For all
$Q\in SO(4)$,
 there exist $A,B \in SU(2)$ such that
$Q= \unor^\dagger (A \otimes B^*)\unor$.
2-1: $\phi$ maps exactly two
elements ($(A,B)$ and $(-A, -B)$)
into one ($\phi(A,B)$).
Homomorphism: $\Phi$
preserves group operations.
\end{theo}

Theorems \ref{claim-isom3}
and \ref{claim-isom4}
are proven in most modern treatises
on  quaternions,
albeit using a different language,
the language of quaternions.
See Version 2 or higher of
Ref.\cite{Paulinesia},
for proofs of
Theorems \ref{claim-isom3}
and \ref{claim-isom4},
given in
the
language
favored here and within the quantum
computing community.

\begin{lemma}\label{claim-re-im-x}
Suppose $X$ is a unitary matrix and
define
$X_R = \frac{X + X^*}{2}$,
$X_I = \frac{X - X^*}{2i}$.
Then $Q=\left(
\begin{array}{cc}
X_R & X_I \\
-X_I & X_R
\end{array} \right)$
is an orthogonal matrix.
Furthermore, $X_R$ and $X_I$ are
real matrices satisfying
$X_R X_R^T + X_I X_I^T =
X_R^T X_R + X_I^T X_I = 1$. Furthermore,
$X_I X_R^T$ and $X_I^T X_R$
are both real, symmetric matrices.
\end{lemma}
\proof
\beq
1=X X^\dagger=(X_R + i X_I)(X_R^T - i X_I^T)
\;,
\eeq
so
\begin{subequations}
\label{eq-w-ortho1}
\beq
X_R X_R^T + X_I X_I^T = 1
\;,
\eeq
and

\beq
X_I X_R^T - X_R X_I^T=0
\;.
\label{eq-w-ortho-sym1}
\eeq
\end{subequations}
From $1=X^\dagger X$ we also get
\begin{subequations}
\label{eq-w-ortho2}
\beq
X_R^T X_R + X_I^T X_I = 1
\;,
\eeq
and

\beq
X_I^T X_R - X_R^T X_I=0
\;.
\label{eq-w-ortho-sym2}
\eeq
\end{subequations}
Note that Eqs.(\ref{eq-w-ortho1})
and Eqs.(\ref{eq-w-ortho2})
are identical except
that
in Eqs.(\ref{eq-w-ortho1}),
the second matrix of
each product is transposed,
whereas in Eqs.(\ref{eq-w-ortho2}),
the first is.
$Q$ is clearly a real matrix, and
Eqs.(\ref{eq-w-ortho1}) imply that
its columns
are orthonormal. Hence $Q$ is orthogonal.
Eq.(\ref{eq-w-ortho-sym1})
(ditto, Eq.(\ref{eq-w-ortho-sym2}))
implies that $ X_I X_R^T$
(ditto, $ X_I^T X_R$) is symmetric.
\qed

The next theorem, due to Eckart and Young,
gives necessary and sufficient
conditions for finding
a pair of unitary matrices $U, V$ that
simultaneously accomplish the SVD
(singular
value decomposition) of
two same-sized but otherwise arbitrary
matrices $A$ and $B$.
The proof reveals that
the problem of finding
simultaneous SVD's
can be reduced to
the simpler problem of
finding simultaneous diagonalizations
of two commuting Hermitian matrices.
The problem of simultaneously diagonalizing
two commuting Hermitian operators
(a.k.a. observables) is well
known to physicists from their study
of Quantum Mechanics.

\begin{theo}\label{claim-ey}
(Eckart-Young)
Suppose $A, B$ are two complex (ditto, real)
rectangular matrices of the same size.
There exist two unitary
(ditto, orthogonal) matrices
$U,V$ such that
$D_1 = U^\dagger A V$ and
$D_2 = U^\dagger B V$ are both
real diagonal matrices
if and only if
$AB^\dagger$ and $A^\dagger B$
are Hermitian (ditto, real symmetric) matrices.
\end{theo}
\proof

($\Rightarrow$)
$AB^\dagger = U D_1 D_2 U^\dagger$
and
$A^\dagger B = V D_1 D_2 V^\dagger$
so they are Hermitian.

($\Leftarrow$)Let

\beq
A' = U_A^\dagger A V_A
= \left(
\begin{array}{cc}
D & 0_2\\
0_3 & 0_4
\end{array}
\right)
\;
\eeq
be a SVD of $A$.
Thus, $U_A, V_A$ are
unitary matrices, $0_2, 0_3, 0_4$
are zero matrices, and
$D$ is a square diagonal matrix
whose diagonal elements are
strictly positive. Let

\beq
B' = U_A^\dagger B V_A
= \left(
\begin{array}{cc}
G & K\\
L & H
\end{array}
\right)
\;,
\eeq
where $D$ and $G$ are square
matrices of the same dimension, $rank(A)$.
Note that

\beq
A'B^{'\dagger}=B' A^{'\dagger}\Rightarrow
\left(
\begin{array}{cc}
D G^\dagger & D L^\dagger\\
0 & 0
\end{array}
\right)=
\left(
\begin{array}{cc}
G D & 0\\
L D & 0
\end{array}
\right)
\;,
\eeq
and

\beq
A^{'\dagger}B'=B^{'\dagger}A' \Rightarrow
\left(
\begin{array}{cc}
D G & D K\\
0 & 0
\end{array}
\right)
=
\left(
\begin{array}{cc}
G^\dagger D  & 0 \\
K^\dagger D & 0
\end{array}
\right)
\;.
\eeq
Therefore,

\beq
L=K=0
\;,
\eeq
and

\beq
D G^\dagger= GD \;,\;\;
D G = G^\dagger D
\;.
\label{eq-commute}
\eeq
When written in index notation,
Eqs.(\ref{eq-commute}) become

\beq
d_i g^*_{ji} = g_{ij}d_j\;,\;\;
d_i g_{ij}= g^*_{ji}d_j
\;,
\label{eq-index1}
\eeq
where the indices range over
$\{1,2,\ldots,rank(A)\}$.
Eqs.(\ref{eq-index1}) imply

\beq
(d_i + d_j)(g^*_{ji} - g_{ij})=0
\;.
\eeq
Since $d_i>0$,
we conclude that $G$
is a  Hermitian matrix.
$D$ is Hermitian too, and, by
virtue of Eq.(\ref{eq-commute}),
$D$ and $G$ commute.
Thus, these two commuting
observables can be diagonalized
simultaneously. Let $P$
be a unitary matrix that accomplishes
this diagonalization:

\beq
D = P^\dagger D P\;,\;\;
D_G = P^\dagger G P
\;.
\eeq
Let

\beq
D_H = U_H^\dagger H V_H
\;
\eeq
be a SVD of $H$.
$D_H$ is a diagonal matrix
with non-negative diagonal entries and
$U_H, V_H$ are unitary matrices.
Now let

\beq
U^\dagger =
\left(
\begin{array}{cc}
P^\dagger & 0\\
0 & U^\dagger_H
\end{array}
\right)
U^\dagger_A
\;,\;\;
V =
V_A
\left(
\begin{array}{cc}
P & 0\\
0 & V_H
\end{array}
\right)
\;.
\label{eq-u-v-defs}
\eeq
The matrices $U$ and $V$
defined by Eq.(\ref{eq-u-v-defs})
can be taken to be the matrices $U$ and $V$ defined
in the statement of the theorem.
\qed

\begin{coro}\label{claim-odo}
If $X$ is a
unitary matrix, then there exist
orthogonal matrices $Q_L$ and $Q_R$ and
a diagonal unitary matrix $e^{i\Theta}$
such that $X= Q_L e^{i\Theta} Q_R^T$.
\end{coro}
\proof
Let $X_R$ and $X_I$ be defined
as in Lemma \ref{claim-re-im-x}.
According to Lemma \ref{claim-re-im-x},
$X_I X_R^T$ and $X_I^T X_R$
are real symmetric matrices, so we can
apply Theorem \ref{claim-ey} with
$A=X_R$ and $B=X_I$. Thus,
there exist orthogonal matrices
$Q_R$ and $Q_L$ such that

\beq
D_R = Q_L^T X_R Q_R\;,\;\;
D_I = Q_L^T X_I Q_R
\;,
\label{eq-dr-di}
\eeq
where $D_R, D_I$ are real diagonal matrices.
Since $X$ is unitary, $D_R + i D_I$ is too.
Thus, we can define a
diagonal unitary matrix $e^{i\Theta}$ by

\beq
e^{i\Theta} = D_R + i D_I
\;.
\label{eq-eit}
\eeq
Combining Eqs.(\ref{eq-dr-di})
and (\ref{eq-eit}) finally yields

\beq
e^{i\Theta} =
Q_L^T X Q_R
\;.
\eeq
\qed

Let
$t=(\theta_1, \theta_2,
\theta_3, \theta_4)\in \RR^4$
and
$\Theta = diag(t)$ so that

\beq
e^{i\Theta} = diag(
e^{i\theta_1},
e^{i\theta_2},
e^{i\theta_3},
e^{i\theta_4})
\;.
\eeq
Let $(k_0, \vec{k})\in \RR^4$.
According to Eq.(\ref{tab-mh-m}),

\beq
 \unor^\dagger\eikks\unor
 =e^{i(k_0 + k_1 \sigzu
 - k_2 \siguz
 + k_3 \sigzz)}
 \;.
 \eeq
If we set

 \beq
 e^{i \Theta}  =\unor^\dagger\eikks\unor
 \;,
 \label{eq-eit-eik}
 \eeq
 then each point
 $(\theta_1, \theta_2, \theta_3, \theta_4)\in \RR^4$
 is mapped in a 1-1
 onto fashion into each point $(k_0, \vec{k})\in \RR^4$.
Using the explicit forms of
$\sigzu, \siguz, \sigzz$, one
 finds that

\beq
\left(
\begin{array}{c}
\theta_0\\
\theta_1\\
\theta_2\\
\theta_3
\end{array}
\right)
= \Gamma
\left(
\begin{array}{c}
k_0\\
k_1\\
k_2\\
k_3
\end{array}
\right)
\;, \;\;
{\rm where}\;\;
\Gamma=
\left(
\begin{array}{cccc}
\pone&\pone&\none&\pone\\
\pone&\pone&\pone&\none\\
\pone&\none&\none&\none\\
\pone&\none&\pone&\pone
\end{array}
\right)
\;.
\eeq
It is easy to check that

\beq
\Gamma^{-1} = \frac{\Gamma^T}{4}
\;.
\eeq

\begin{coro}
\label{claim-kak1}
(KAK1)
If $X\in U(4)$, then
$X = (A_1\otimes A_0)\eikks (B_1 \otimes B_0)$,
where $A_1, A_0, B_1, B_0\in SU(2)$
and $(k_0, \vec{k})\in \RR^4$.
\end{coro}
\proof
Let
\beq
X' = \unor^\dagger X \unor
\;.
\eeq
$X'$ is a unitary matrix, so,
according to Collorary \ref{claim-odo},
we can find orthogonal matrices $Q_L, Q_R$
and a diagonal unitary matrix
$e^{i \Theta}$ such that

\beq
X' = Q_L e^{i \Theta} Q_R^T
\;.
\eeq
According to Theorem \ref{claim-isom4},
we can find $A_1, A_0, B_1, B_0\in SU(2)$
such that

\beq
\unor Q_L \unor^\dagger = A_1\otimes A_0
\;,
\eeq
and

\beq
\unor Q_R \unor^\dagger = B_1\otimes B_0
\;.
\eeq
As in Eq.(\ref{eq-eit-eik}), set

\beq
\unor e^{i \Theta} \unor^\dagger =\eikks
\;.
\eeq
It follows that

\beq
X = (A_1\otimes A_0)\eikks (B_1 \otimes B_0)
\;.
\eeq
\qed

\section{Canonical Class Vector}

In this section we discuss how
KAK1 partitions
$SU(4)$ into disjoint classes
characterized by a 3d real vector $\vec{k}$.

We will say that $U,V\in SU(4)$
{\bf are equivalent up to local
operations} and write $U\sim V$
if $U =(R_1 \otimes R_0)
V (S_1 \otimes S_0)$ where
$R_1, R_0, S_1, S_0 \in U(2)$.
It is easy to prove that
$\sim$ is an equivalence relation.
Hence, it partitions $SU(4)$
into
disjoint subsets (i.e., equivalence classes).
If $X\in SU(4)$ and
$\vec{k}\in \RR^3$ are related
as in Collorary
\ref{claim-kak1}, then
$X\sim \eiks$.
Henceforth, we will call this
$\vec{k}$
 a {\bf class vector}
of $X$.
We will say that $\vec{k'}$ and
$\vec{k}$ are {\bf equivalent
class vectors} and write
$\vec{k'}\sim \vec{k}$
if
$\eikps \sim
\eiks$.

Note that the following 3
operations map a class
vector into another class
vector of the same class; i.e.,
the operations are class-preserving.

\begin{enumerate}
\item (Shift)
Suppose we shift
$\vec{k}$ by plus or minus $\frac{\pi}{2}$
along any one of its 3 components.
For example, a positive, $\frac{\pi}{2}$,  X-shift would map
\beq
(k_x, k_y, k_z)\mapsto
(k_x + \frac{\pi}{2}, k_y, k_z)
\;.
\eeq
This operation preserves $\vec{k}$'s
class because

\beq
e^{i[(k_x + \frac{\pi}{2})\sigxx +
k_y \sigyy + k_z \sigzz]}
=
e^{i\frac{\pi}{2}\sigxx}\eiks\\
=
i\sigxx \eiks
\;.
\eeq

\item (Reverse)
Suppose we reverse the
sign of any two components of $\vec{k}$.
For example, an XY-reversal would map

\beq
(k_x, k_y, k_z)\mapsto
(-k_x, -k_y, k_z)
\;.
\eeq
This operation preserves $\vec{k}$'s
class because

\beq
\sigzu \eiks \sigzu
=
e^{i(-k_x, -k_y, k_z)\cdot \vec{\Sigma}}
\;.
\eeq

\item (Swap)
Suppose we swap
any two components of $\vec{k}$.
For example, an XY-swap would map

\beq
(k_x, k_y, k_z)\mapsto
(k_y, k_x, k_z)
\;.
\eeq
This operation preserves $\vec{k}$'s
class because

\beq
e^{-i\frac{\pi}{4}(\sigzu + \siguz)}
\eiks
e^{i\frac{\pi}{4}(\sigzu + \siguz)}
=
e^{i(k_y, k_x, k_z)\cdot \vec{\Sigma}}
\;.
\eeq
\end{enumerate}

Define ${\cal K}$
as the set of points $\vec{k}\in \RR^3$
such that

\begin{figure}[h]
    \begin{center}
    \epsfig{file=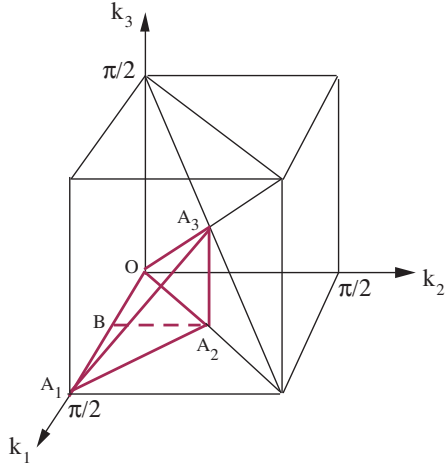, height=2.5in}
    \caption{
    The set ${\cal K}$ of canonical class vectors
    equals the set of points in the
    tetrahedral region $OA_1A_2A_3$
    (including all the interior and
    the surface points
    except that only the half $OBA_2$
    of the base $OA_1A_2$
    is included).}
    \label{fig-canon-vec}
    \end{center}
\end{figure}

\begin{enumerate}
\item
$\frac{\pi}{2}
>k_x \geq k_y \geq k_z \geq 0$
\item
$k_x + k_y \leq \frac{\pi}{2}$
\item
If $k_z =0$, then $k_x \leq \frac{\pi}{4}$.
\end{enumerate}
${\cal K}$ is contained within
the tetrahedral region $OA_1A_2A_3$
of Fig.\ref{fig-canon-vec}.

The 3 class-preserving operations
given above
generate a group $\ul{W}$.
Given any class vector
$\vec{k}\in \RR^3$,
it is always possible to find
an operation $G\in \ul{W}$
such that $G(\vec{k})\in{\cal K}$.
Indeed, here is an algorithm,
(implemented in the accompanying
Octave software)
that finds $G(\vec{k})\in{\cal K}$
for any $\vec{k}\in \RR^3$:

\begin{enumerate}
\item
Make $k_x \in [0 \frac{\pi}{2})$
by shifting $k_x$
repeatedly by $\frac{\pi}{2}$.
In the same way, shift
$k_y$ and $k_z$  into $[0 \frac{\pi}{2})$.
\item
Make
$k_x\geq k_y \geq k_z$
by swapping the components of $\vec{k}$.
\item
Perform this step iff at this point
$k_x + k_y > \frac{\pi}{2}$.
Transform $\vec{k}$ into
$(\frac{\pi}{2}-k_y,
\frac{\pi}{2} - k_x, k_z)$
(This can be
achieved by applying an XY-swap,
 XY-reverse, X-shift and
Y-shift, in that order).
At this point, $k_x\geq k_y$, but
$k_z$ may be larger than $k_y$ or $k_x$,
so finish this step by swapping
coordinates until $k_x\geq k_y \geq k_z$ again.
\item
Perform this step iff at this point
$k_z=0$ and $k_x> \frac{\pi}{4}$.
Transform $\vec{k}=(k_x, k_y, 0)$ into
$(\frac{\pi}{2} - k_x, k_y , 0)$
(This can be achieved by
applying an XZ-reverse and an X-shift,
in that order).
\end{enumerate}

We can find a subset $S$
of $\RR^3$
such that
every equivalence class of $SU(4)$ is
represented by one and only one
point $\vec{k}$ of $S$.
In fact, ${\cal K}$
defined above is one such $S$.
We will refer to the elements of ${\cal K}$
as {\bf canonical class vectors}.

We end this section
by finding the canonical class vectors
of some simple 2-qubit operations.

\begin{enumerate}
\item (CNOT):
$CNOT(1\rarrow 0)$ is defined by

\beq
CNOT(1\rarrow 0)= \sigx(0)^{n(1)}=
\left(
\begin{array}{cc}
1 & 0\\
0 & \sigx
\end{array}
\right)
\;.
\eeq
Since $n=\frac{1}{2}(1-\sigz)$,
$n_X=\frac{1}{2}(1-\sigx)$,
and $\sigx=(-1)^{n_X}=e^{i\pi n_X}$,

\begin{subequations}
\begin{eqnarray}
\sigx(0)^{n(1)} &=& (-1)^{n_X(0) n(1)}\\
&=&e^{i \frac{\pi}{4}(1-\sigux)(1-\sigzu)}\\
&=&
e^{i \frac{\pi}{4}(1-\sigzu -\sigux)}
e^{i \frac{\pi}{4}\sigzx}\\
&=&
e^{i \frac{\pi}{4}(1-\sigzu -\sigux)}
e^{i\frac{\pi}{4}\sigyu}
e^{i \frac{\pi}{4}\sigxx}
e^{-i\frac{\pi}{4}\sigyu}\\
&\sim & e^{i\frac{\pi}{4}\sigxx}
\;.
\end{eqnarray}
\end{subequations}
Therefore, the canonical class vector
of CNOT is $(\frac{\pi}{4},0,0)$,
which corresponds to the point
$B$ in Fig.\ref{fig-canon-vec}.

\item ($\sqrt{CNOT}$) From the
math just performed for CNOT,
it is clear that

\beq
\sqrt{\sigx(0)^{n(1)}}=
e^{i\frac{\pi}{8}(1-\sigux)(1-\sigzu)}
\sim e^{i\frac{\pi}{8}\sigxx}
\;.
\eeq
Therefore, the canonical class vector
of $\sqrt{CNOT}$ is $(\frac{\pi}{8},0,0)$,
which corresponds to the midpoint
of the segment
$OB$ in Fig.\ref{fig-canon-vec}.

\item (Exchanger, a.k.a. Swapper)
As usual, the Exchanger is defined by

\beq
E=
\left(
\begin{array}{cccc}
1&0&0&0\\
0&0&1&0\\
0&1&0&0\\
0&0&0&1
\end{array}
\right)
\;.
\eeq
(Note that $\det(E)=-1$).
Using Eqs.(\ref{eq-com-sigxx-sigyy}),
it is easy to show that

\beq
E
=e^{-i \frac{\pi}{4}} e^{i\frac{\pi}{4}
(\sigxx+\sigyy+\sigzz)}
\;.
\eeq
Therefore, the canonical class vector
of $e^{i \frac{\pi}{4}}E$ is
$(\frac{\pi}{4},\frac{\pi}{4},\frac{\pi}{4})$,
which corresponds to the apex
$A_3$ of the tetrahedron
in Fig.\ref{fig-canon-vec}.

\end{enumerate}

\section{Software}

A collection of Octave/Matlab
m-files that implement the algorithms
in this
paper,
can be found at ArXiv (under the
``source" for this paper),
and at my website (www.ar-tiste.com).
These m-files have only
been tested on Octave, but they
should run on Matlab with few or
no modifications.
A file called ``m-fun-index.html"
that accompanies the m-files lists
each function and its purpose.

\end{document}